\documentclass[a4paper,10pt,twoside]{cpc-hepnp}

\usepackage{multicol}
\usepackage{graphicx}
\usepackage{booktabs}
\usepackage{amssymb,bm,mathrsfs,bbm,amscd}
\usepackage[tbtags]{amsmath}
\usepackage{lastpage}

\usepackage{subfigure}
\usepackage{url}
\def\DAF{DA$\Phi$NE} 
\def\ifm#1{\relax\ifmmode#1\else$#1$\fi}
  
\def\epm{\ifm{e^+e^-}} 
\def\dif{\hbox{d}}
\newcommand{\dd}{{\rm d}\,}

\begin{document}

\fancyhead[co]{\footnotesize KLOE coll.: Measurement of the pion form factor for $M_{\pi\pi}^2$ between 0.1 and 0.85 GeV$^2$}


\title{Measurement of the pion form factor for $M_{\pi\pi}^2$ between 0.1 and 0.85 GeV$^2$ with 
the KLOE detector}

\author{KLOE collaboration\thanks{
F.~Ambrosino, 
A.~Antonelli,  
M.~Antonelli,
F.~Archilli,  
P.~Beltrame,  
G.~Bencivenni,  
C.~Bini,  
C.~Bloise,  
S.~Bocchetta,
F.~Bossi,
P.~Branchini,  
G.~Capon,  
T.~Capussela,   
F.~Ceradini, 
P.~Ciambrone,
E.~De~Lucia, 
A.~De~Santis,  
P.~De~Simone,  
G.~De~Zorzi,  
A.~Denig, 
A.~Di~Domenico, 
C.~Di~Donato, 
B.~Di~Micco, 
M.~Dreucci,   
G.~Felici, 
S.~Fiore, 
P.~Franzini, 
C.~Gatti,       
P.~Gauzzi, 
S.~Giovannella, 
E.~Graziani, 
M.~Jacewicz,
W.~Kluge,         
J.~Lee-Franzini,                    
M.~Martini,
P.~Massarotti,
S.~Meola,   
S.~Miscetti,  
M.~Moulson, 
S.~M\"uller, 
F.~Murtas,  
M.~Napolitano, 
F.~Nguyen, 
M.~Palutan,            
A.~Passeri,   
V.~Patera,
P.~Santangelo, 
B.~Sciascia, 
T.~Spadaro,   
L.~Tortora,                      
P.~Valente, 
G.~Venanzoni,     
R.~Versaci,
G.~Xu}
\\
presented by
Stefan E. M\"uller
$^{1)}$\email{muellers@kph.uni-mainz.de}%
}
\maketitle

\address{%
1~Institut f\"ur Kernphysik, Johannes Gutenberg-Universit\"at, Johann-Joachim-Becher-Weg 45, 55128 Mainz, Germany\\
}

\begin{abstract}
The KLOE experiment at the $\phi$-factory DA$\Phi$NE has measured the
pion form factor in the range between $0.1 < M_{\pi\pi}^2 <
0.85$ GeV$^2$ using events taken at $\sqrt{s}= 1$ GeV with a photon emitted 
at large polar angles in the initial state. This measurement extends the $M_{\pi\pi}^2$ region 
covered by KLOE ISR measurements of the pion form factor down to the two pion 
production threshold. The value obtained in this measurement of the dipion 
contribution to the muon anomalous magnetic moment of $\Delta a_\mu^{\pi\pi} = 
(478.5\pm2.0_\mathrm{stat}\pm4.8_\mathrm{syst}\pm2.9_\mathrm{theo})\cdot
10^{-10}$ further confirms the discrepancy between the
Standard Model evaluation for $a_\mu$ and the experimental value
measured by the (g-2) collaboration at BNL.  
\end{abstract}



\begin{multicols}{2}

\section{Introduction}
The anomalous magnetic moment of the muon, $a_\mu$, is one of the best known quantities in particle physics. Recent theoretical 
evaluations~\cite{Jegerlehner:2009ry,Teubner:2009,Davier:2009} 
find a discrepancy
of 3 - 4 standard deviations from the value obtained from the 
g-2 experiment at Brookhaven~\cite{g-2:2006,Roberts:2009}. A large part of the 
uncertainty on the
theoretical estimates comes from the leading order hadronic
contribution $a_\mu^{\mathrm{had,lo}}$, which at
low energies is not calculable by perturbative QCD, but has to be
evaluated with a dispersion integral using measured 
hadronic cross sections. The use of initial state 
radiation (ISR) has opened a new way to obtain these cross sections at
particle factories operating at fixed energies~\cite{actis:2009gg}. 
The region below 1 GeV, which 
is accessible with the KLOE experiment in 
Frascati, is dominated by the $\pi^+\pi^-$ final state and contributes with 
$\sim 70\%$ to $a_\mu^{\mathrm{had,lo}}$, and $\sim 60\%$ to its
uncertainty. Therefore, improved precision in the $\pi\pi$ cross
section would result in a reduction of the uncertainty on the leading
order hadronic contribution to $a_\mu$, and in turn improve the
Standard Model prediction for $a_\mu$.

\section{Measurement of $\sigma_{\pi\pi}$}
The measurement has been performed with the KLOE detector at the
\DAF\, $e^+e^-$ collider in Frascati. \DAF\, is a $\phi$-factory that usually operates at $\sqrt{s}\simeq M_\phi$, and has delivered ca. 2.5
fb$^{-1}$ of data to the KLOE experiment up to the year 2006, from which KLOE has reported two measurements of the 
$\pi\pi$ cross section between 0.35 and 0.95 GeV$^2$~\cite{plb606,plb670}. In addition,
about 250 pb$^{-1}$ of data have been collected at $\sqrt{s}\simeq 1$
GeV, 20 MeV below the $\phi$ resonance, from which the new results were 
obtained. Running below the $\phi$ resonance diminishes the backgrounds from the copious $\phi$ decay products, including scalar mesons.
As \DAF\, was designed to operate at a fixed energy around $M_\phi$, the
differential cross section
$\dif \sigma(\epm\to\pi^+\pi^-+\gamma_{\mathrm{ISR}})/ \dif M_{\pi\pi}^2$
is measured, and the total cross section
$\sigma_{\pi\pi}\equiv\sigma_{e^+ e^-\to\pi^+\pi^-}$ is evaluated
using the formula~\cite{Binner99}:  
\begin{equation}
s\cdot \frac{\dif\sigma_{\pi\pi\gamma_\mathrm{ISR}}}
{\dif M_{\pi\pi}^2} = \sigma_{\pi\pi}
(M_{\pi\pi}^2)~ H(M_{\pi\pi}^2,s)~,
\label{eq:1}
\end{equation}
in which $s$ is the squared $e^+e^-$ center of mass energy, and $H$ is a  
radiator function obtained from theory describing the photon emission in the initial state. Final State Radiation (FSR)
terms are neglected in Eq.~\ref{eq:1}, but are taken into account properly in
the analysis.
\begin{center}
\includegraphics[width=15.pc]{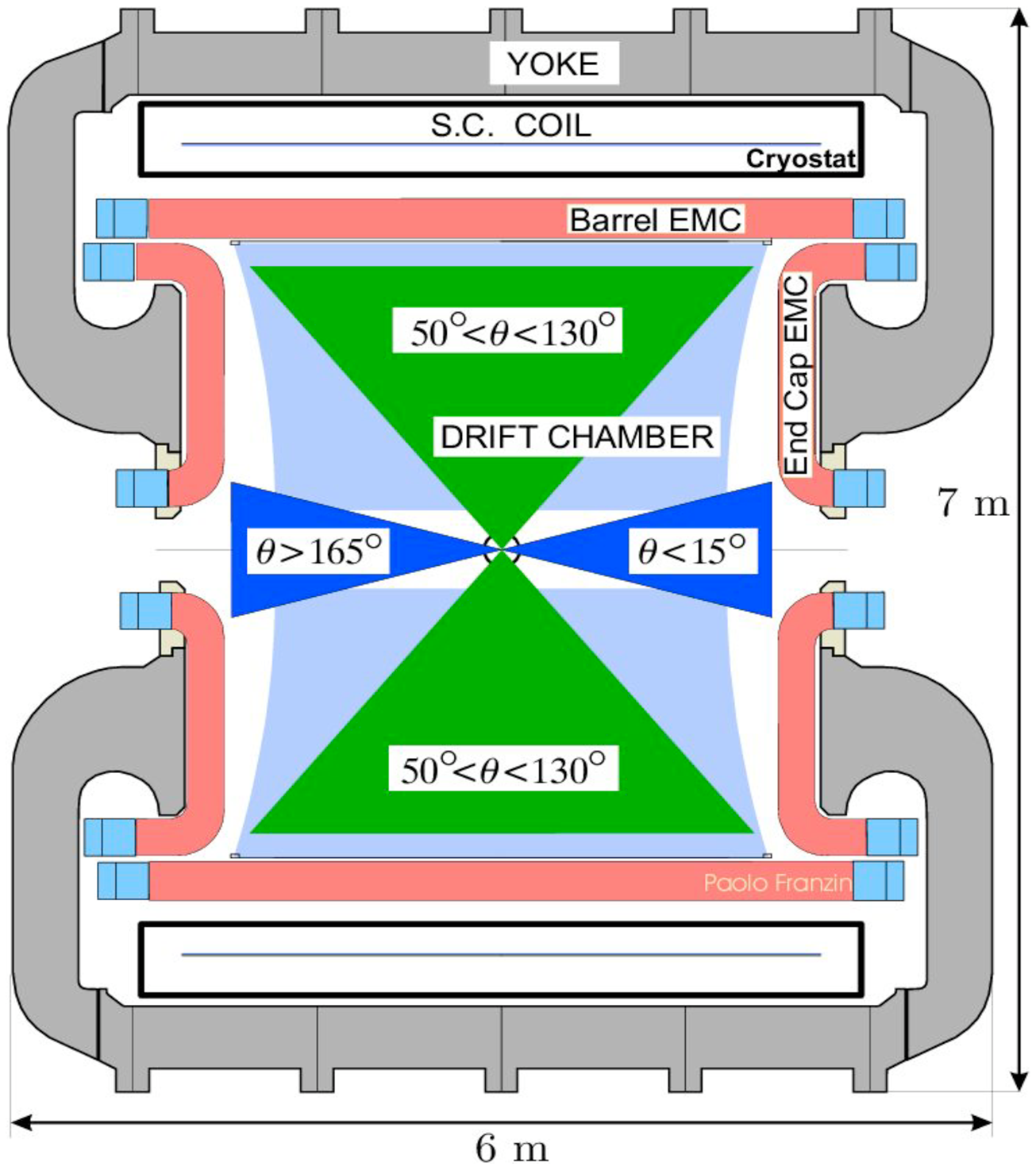}
\figcaption{\label{fig:1} Schematic view of the KLOE detector with selection regions.}
\end{center}
The KLOE detector (Fig.~\ref{fig:1}) consists of a high
resolution drift chamber ($\sigma_{p} / p \leq 0.4\%$)~\cite{KLOE:DC} and an 
electromagnetic calorimeter with excellent time ($\sigma_t\sim
54 ~\mathrm{ps}/\sqrt{E~[\mathrm{GeV}]}$ $\oplus100~\mathrm{ps}$) and good energy 
($\sigma_E/E\sim 5.7\%/\sqrt
{E~[\mathrm{GeV}]}$) resolution~\cite{KLOE:EMC}. 
\subsection{Event selection}
The previous KLOE analyses~\cite{plb606,plb670} used selection cuts in 
which photons are emitted
within a cone of $\theta_\gamma<15^\circ$ around the 
beamline (narrow cones in Fig.~\ref{fig:1}) and the two charged pion tracks 
 have $50^\circ<\theta_\pi<130^\circ$ (wide cones in Fig.~\ref{fig:1}). 
In this configuration,  the photon is
not explicitly detected, its direction
is reconstructed from the tracks' momenta
by closing kinematics: $\vec{p}_\gamma\simeq\vec{p}_\mathrm{miss}= -(\vec{p}_{\pi^+}
+\vec{p}_{\pi^-})$. While these cuts guarantee a high
  statistics for ISR signal events, and a reduced contamination from the resonant process $e^+e^-\to
  \phi\to\pi^+\pi^-\pi^0$ in which the $\pi^0$ mimics the missing
  momentum of the photon(s) and from the final state 
radiation process $e^+e^-\to \pi^+\pi^-\gamma_\mathrm{FSR}$,  a highly 
energetic 
photon emitted at small angle forces the pions also to be at small
  angles (and thus outside the selection cuts), resulting in a 
kinematical suppression of events with $M^2_{\pi\pi}< 0.35$
  GeV$^2$. To access the two pion threshold, a new analysis is performed requiring events that are 
selected to have a photon at large polar angles between 
$50^\circ<\theta_\gamma<130^\circ$ (wide cones in Fig.~\ref{fig:1}), 
in the same angular region as the pions to be included. The drawback using such acceptance 
cuts is a reduction in statistics of about a factor 5, as well as an increase
of events with final state radiation and from $\phi$ radiative decays
compared to 
the small angle photon acceptance criterion. The uncertainty on the model 
dependence of the $\phi$ radiative decays to the scalars $f_0(980)$ and 
$f_0(600)$ together with $\phi\to\rho\pi\to(\pi\gamma)\pi$ has a strong impact on the measurement~\cite{Leone:2007zz}. As an obvious way out of this dilemma, the present analysis  
uses the data taken by the KLOE experiment in 2006 at a value of $\sqrt{s}=1$ GeV, 
about 5 $\Gamma_\phi$ outside the narrow peak of the
$\phi$ resonance ($\Gamma_\phi = 4.26\pm0.04$
MeV~\cite{Amsler:2008zzb}). This reduces the effect
due to contributions from $f_0\gamma$ and $\varrho\pi$ decays of the
$\phi$-meson to within $\pm 1\%$. 
\begin{center}
\includegraphics[width=14.pc]{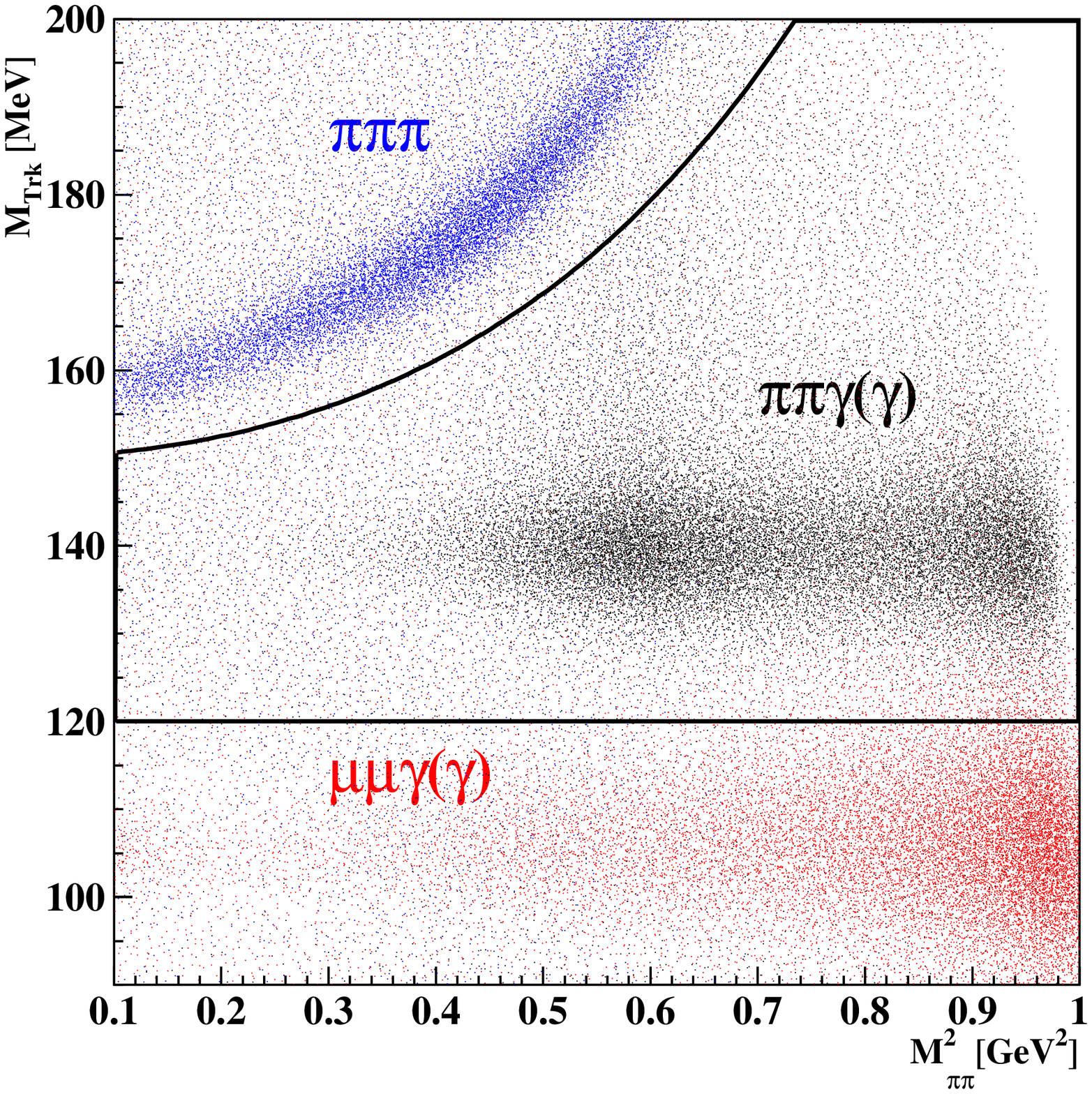}
\figcaption{\label{fig:2} MC simulation of $M_\mathrm{trk}$ vs. $M_{\pi\pi}^2$.  $\pi^+\pi^-\gamma$ and $\mu^+\mu^-\gamma$  events are located around $m_\pi$ and $m_\mu$ respectively, while $\pi^+\pi^-\pi^0$ events occupy a region in the upper left of the plot. The black lines represent the cuts used in the analysis.}
\end{center}
Contaminations from the processes $\phi\to\pi^+\pi^-\pi^0$
and $\epm\to\mu^+\mu^-\gamma$ are rejected by cuts in 
the kinematical variables {\it trackmass}\footnote{The trackmass is defined using conservation of 4-momentum under the
hypothesis that the final state consists of two charged particles with
equal mass $M_\mathrm{trk}$ and one photon.} and $\Omega$\footnote{$\Omega$ is the three-dimensional angle between the direction of the selected photon 
and the missing momentum.}
(see Fig.~\ref{fig:2} and Fig.~\ref{fig:3}). 
A particle ID estimator based
on calorimeter information and time-of-flight is used to efficiently suppress 
the high rate of radiative Bhabhas.
The radiative differential cross section is then obtained subtracting
the re\-si\-dual background events, $N_{bkg}$, dividing by
the selection efficiencies, $\varepsilon_{sel}(M_{\pi\pi}^2)$,
and the integrated luminosity using the formula
\begin{equation}
\frac{\dd\sigma_{\pi\pi\gamma}}
{\dd M_{\pi\pi}^2} = \frac{N_{obs}-N_{bkg}}
{\Delta M_{\pi\pi}^2}\, \frac{1}{\varepsilon_{sel}(M_{\pi\pi}^2)~ \mathcal{L}}~ ,
\label{eq:2}
\end{equation}
where the observed events are selected in bins of
$\Delta M_{\pi\pi}^2=0.01$ GeV$^2$. 
The residual background content is found by fitting the
$M_{\mathrm trk}$ spectrum of the selected data sample with a superposition
of Monte Carlo distributions describing the signal and
background sources. The fit parameters 
are the fractional normalization factors for these Monte Carlo distributions,
obtained in intervals of $M^2_{\pi\pi}$.     
\begin{center}
\includegraphics[width=14.pc]{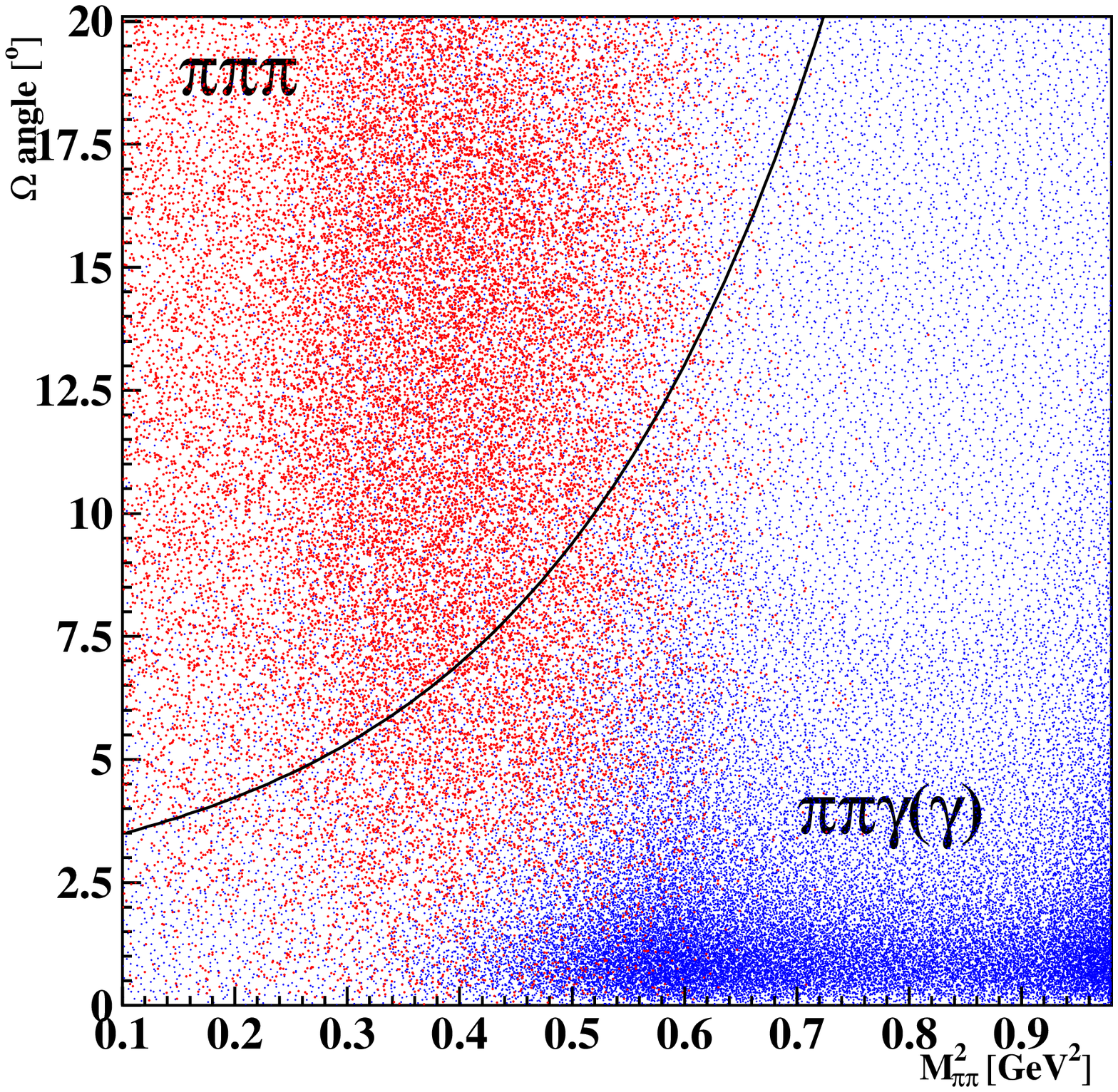}
\figcaption{\label{fig:3} MC simulation of $\Omega$-angle vs. $M_{\pi\pi}^2$. $\pi^+\pi^-\gamma$-events (blue) are distributed at small values of $\Omega$, while the $\pi^+\pi^-\pi^0$ events (red) occupy the region below 0.5 GeV$^2$ at larger values of $\Omega$. The black line represents the cut used in the analysis.}
\end{center}
\subsection{Luminosity}
The absolute normalization of the data sample is performed by
measuring Bhabha events at large angles ($55^\circ<\theta<125^\circ$),
with an effective cross section of $\sigma_\mathrm{Bhabha}\simeq 430$ nb. To obtain the
integrated luminosity, $\mathcal{L}$, the observed number
of Bhabha events is divided by the
effective cross section evaluated by the Monte Carlo generator
\texttt{Babayaga@NLO}~\cite{CarloniCalame:2000pz,Balossini:2006wc},
which includes QED radiative corrections
with the parton shower algorithm, and which has been interfaced with
the KLOE detector simulation. A detailed description of the KLOE luminosity measurement can be found
in~\cite{Ambrosino:2006te}.
\subsection{Radiative corrections}
The radiator function $H$ used to get $\sigma_{\pi\pi}$ in
Eq.~\ref{eq:1} is obtained from the \texttt{PHOKHARA} Monte Carlo generator,
which calculates the complete next-to-leading order ISR
effects~\cite{Czyz:2003}. In addition, the cross section is corrected
for the vacuum polarisation~\cite{Jeger_alpha} (running of $\alpha_{\mathrm em}$), 
and the
shift between the measured value of $M^2_{\pi\pi}$ and the squared 
virtual photon
mass $M^2_{\gamma^*}\equiv(M^0_{\pi\pi})^2$ for events with photons from final state
radiation. Again the \texttt{PHOKHARA} generator, 
which includes FSR effects in the pointlike-pions approximation, 
is used to estimate the latter~\cite{Czyz:2005}, and a matrix relating 
$M^2_{\pi\pi}$ to 
$M^2_{\gamma^*}$ by giving the probability for an event in a bin of $M^2_{\pi\pi}$ to end up in a bin of $M^2_{\gamma^*}$ is used to correct the spectrum.

\subsection{Results}
Using Eq.~\ref{eq:1} and Eq.~\ref{eq:2}, one obtains the two-pion cross
section $\sigma_{\pi\pi}$. The squared modulus of the pion form factor
$|F_\pi|^2$ can then be derived using the relation\footnote{In
  addition, the choice of radiative corrections applied to
  $\sigma_{\pi\pi}$ and $|F_\pi|^2$ may differ betwen the two. We
  adopt the definition used in~\cite{CMD2,CMD2_2,SND}, in which
  $\sigma_{\pi\pi}$ is inclusive with respect to final state
  radiation, and undressed from vacuum polarisation effects; while
  $|F_\pi|^2$ contains vacuum polarisation effects and final state
  radiation is removed.}
\begin{equation} 
|F_\pi(s')|^2=\frac{3}{\pi}\frac{s'}{\alpha_{\rm em}^2
  \beta_\pi^3}\sigma_{\pi\pi}(s')~, 
\label{eq:3}
\end{equation}
where $s'=(M^0_{\pi\pi})^2$ is the squared momentum transferred by the 
virtual photon and $\beta_\pi = \sqrt{1-
\frac{4m_\pi^2}{s'}}$ is the pion velocity.

Fig.~\ref{fig:4} shows $|F_\pi|^2$ as a function of
$(M^0_{\pi\pi})^2$ for the new KLOE09 measurement and the previous
KLOE publication, KLOE08. As can be seen from Fig.~\ref{fig:5}, both
measurements are in very good agreement, especially above 0.5 GeV$^2$.

\begin{center}
\includegraphics[width=18.pc]{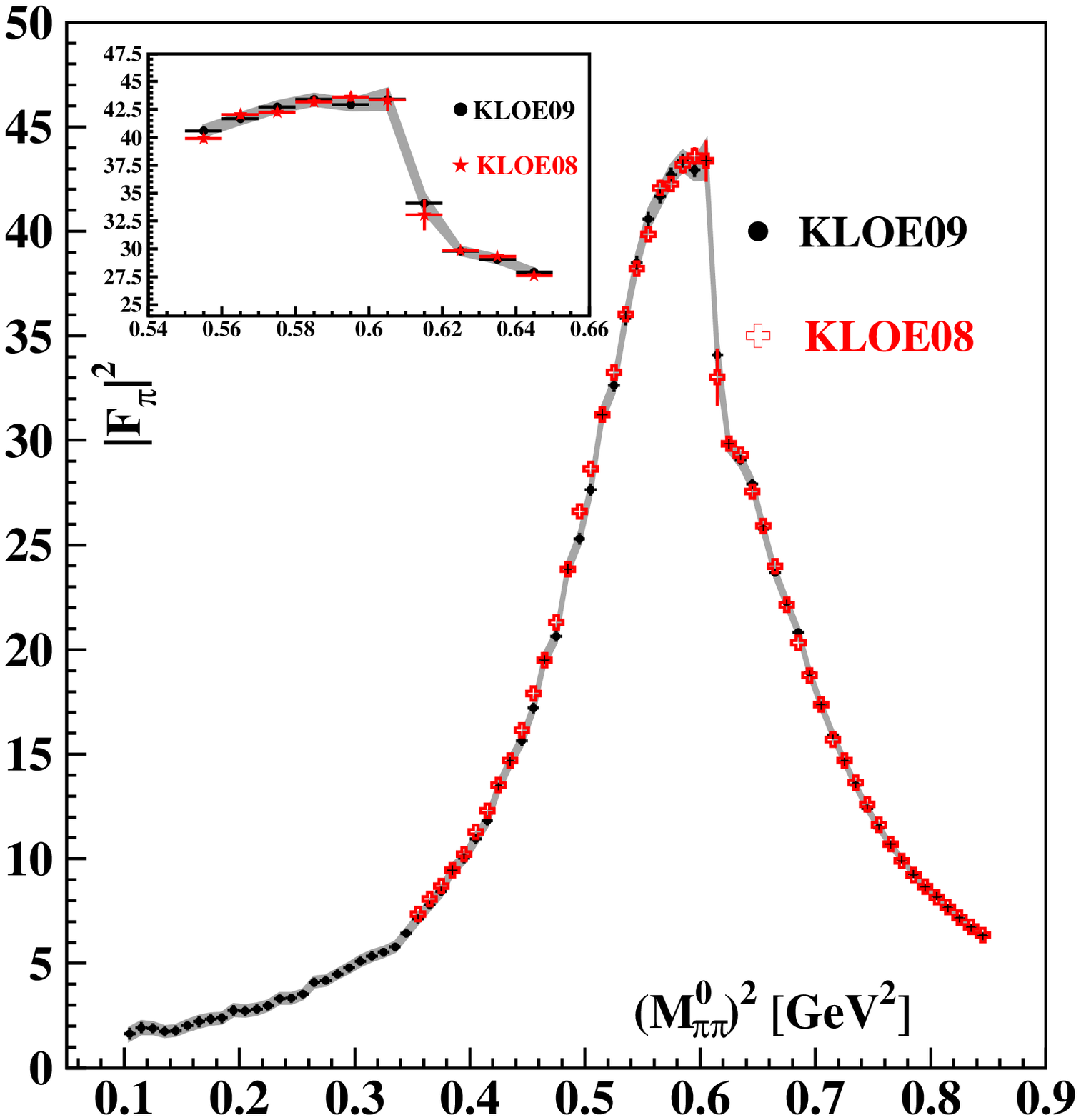}
\figcaption{\label{fig:4} Pion form factor $|F_\pi^2|$ obtained in the present (KLOE09) and the previous (KLOE08) analysis. KLOE09 data points have statistical error attached, the grey band gives the statistical and systematic uncertainty (added in quadrature). Errors on KLOE08 points contain the combined statistical and systematic uncertainty.} 
\end{center}
\begin{center}
\includegraphics[width=18.pc]{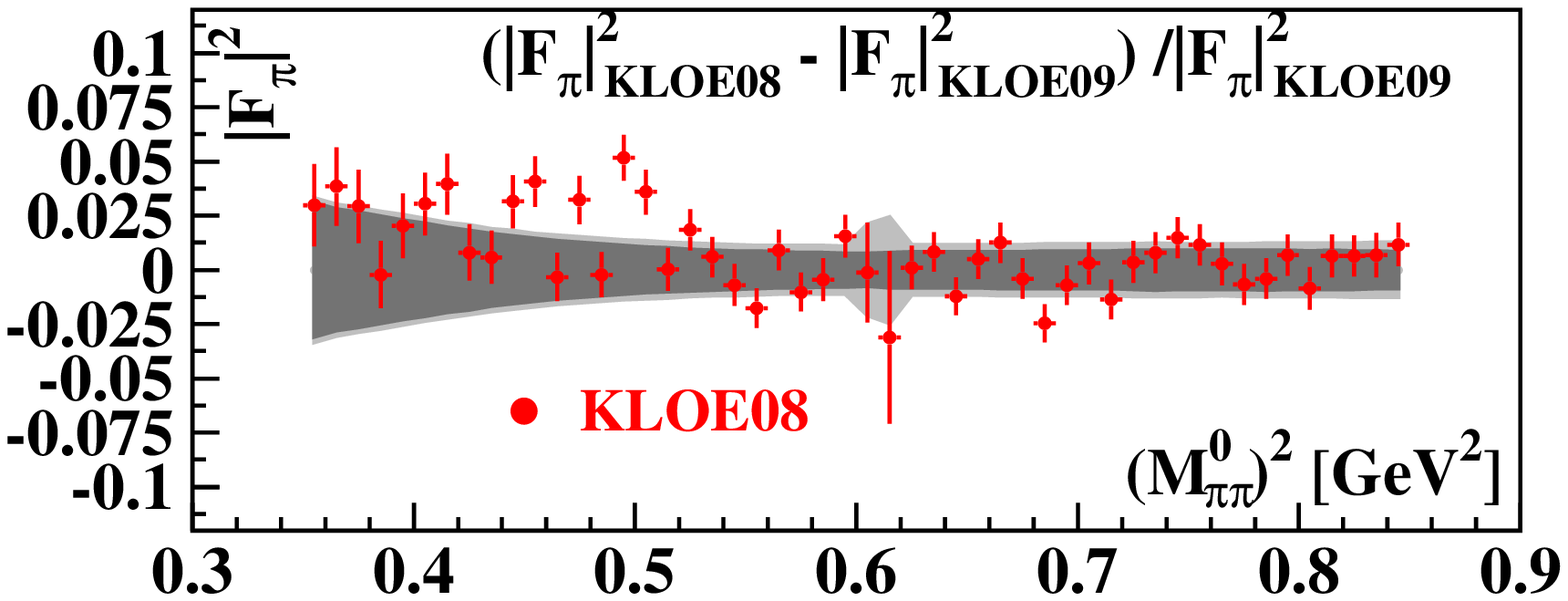}
\figcaption{\label{fig:5} Fractional difference between $|F_\pi^2|$ from the KLOE08 and the KLOE09 analysis. The band in dark grey represents the statistical error of the KLOE09 result, the band in lighter grey gives the statistical and systematic uncertainty (added in quadrature) for the KLOE09 result. Errors on KLOE08 points contain the combined statistical and systematic uncertainty.}
\end{center}

The cross section corrected for the running of $\alpha_{em}$ and
 inclusive of FSR, $\sigma_{\pi\pi(\gamma)}^{bare}$,
is used to determine $\Delta a_\mu^{\pi\pi}$ via a dispersion integral:
\begin{equation}
\label{eq:4}
a_\mu^{\pi\pi} = 
\frac{1}{4\pi^3}\int_{s_{min}}^{s_{max}}\dd s'~
\sigma_{\pi\pi(\gamma)}^{bare}(s')\,K(s')~ ,
\end{equation}
where the lower and upper bounds are
$s_{min}=0.10$ GeV$^2$ and $s_{max}=0.85$ GeV$^2$ in the present analysis, and the kernel function $K(s)$ is described in~\cite{Brodsky:67}. We obtain a value of
\begin{eqnarray}
\lefteqn{\Delta a_\mu^{\pi\pi}(0.1-0.85\; {\rm GeV^2}) = {} }\nonumber\\
 & & {}(478.5 \pm 2.0_{\rm stat} \pm 4.8_{\rm exp} \pm 2.9_{\rm theo}) \cdot 10^{-10}.
  \label{eqn:amupipi_pop}
\end{eqnarray}
\begin{center}
\begin{tabular}{||l|c||}
\hline
Reconstruction Filter & negligible\\
Background subtraction & 0.5 \% \\
f$_0+\rho\pi$ bkg. & 0.4 \% \\
$\Omega$ cut & 0.2 \% \\
Trackmass cut & 0.5 \% \\
$\pi$/e-ID & negligible\\
Tracking & 0.3 \% \\
Trigger & 0.2 \% \\
Acceptance & 0.4 \% \\
Unfolding & negligible \\
Software Trigger (L3) & 0.1 \% \\
Luminosity ($0.1_{th}\oplus 0.3_{exp}$)\% & 0.3 \% \\
\hline
Total exp. systematics & 1.0 \% \\
\hline
\hline
FSR resummation & 0.3 \% \\
Vacuum Polarization &  0.1 \% \\
Rad. function $H$   & 0.5 \% \\
\hline
Total theory systematics & 0.6 \% \\
\hline
\end{tabular}
\vspace{0.3cm}
\tabcaption{\label{tab:2} List of systematic errors on the $\Delta a_\mu^{\pi\pi}$ evaluation.}
\end{center}
The evaluation of $\Delta a_\mu^{\pi\pi}$  in the range between 0.35 and 0.85 GeV$^2$ allows to compare the result obtained in this new analysis with the previously published result by KLOE~\cite{plb670}:
\begin{center}
  \renewcommand{\arraystretch}{1.5}
  \begin{tabular}{l|c}
    KLOE Analysis & $\Delta a_\mu^{\pi\pi} (0.35 - 0.85\; {\rm GeV^2}) \times 10^{-10} $ \\
    \hline \hline
    KLOE09 & $376.6 \pm 0.9_{\rm stat} \pm 2.4_{\rm exp} \pm 2.1_{\rm theo}$ \\
    KLOE08 & $379.6 \pm 0.4_{\rm stat} \pm 2.4_{\rm exp} \pm 2.2_{\rm theo}$ \\
    \hline
  \end{tabular}
\end{center}
The two values are in good agreement.
\begin{center}
\includegraphics[width=18.pc]{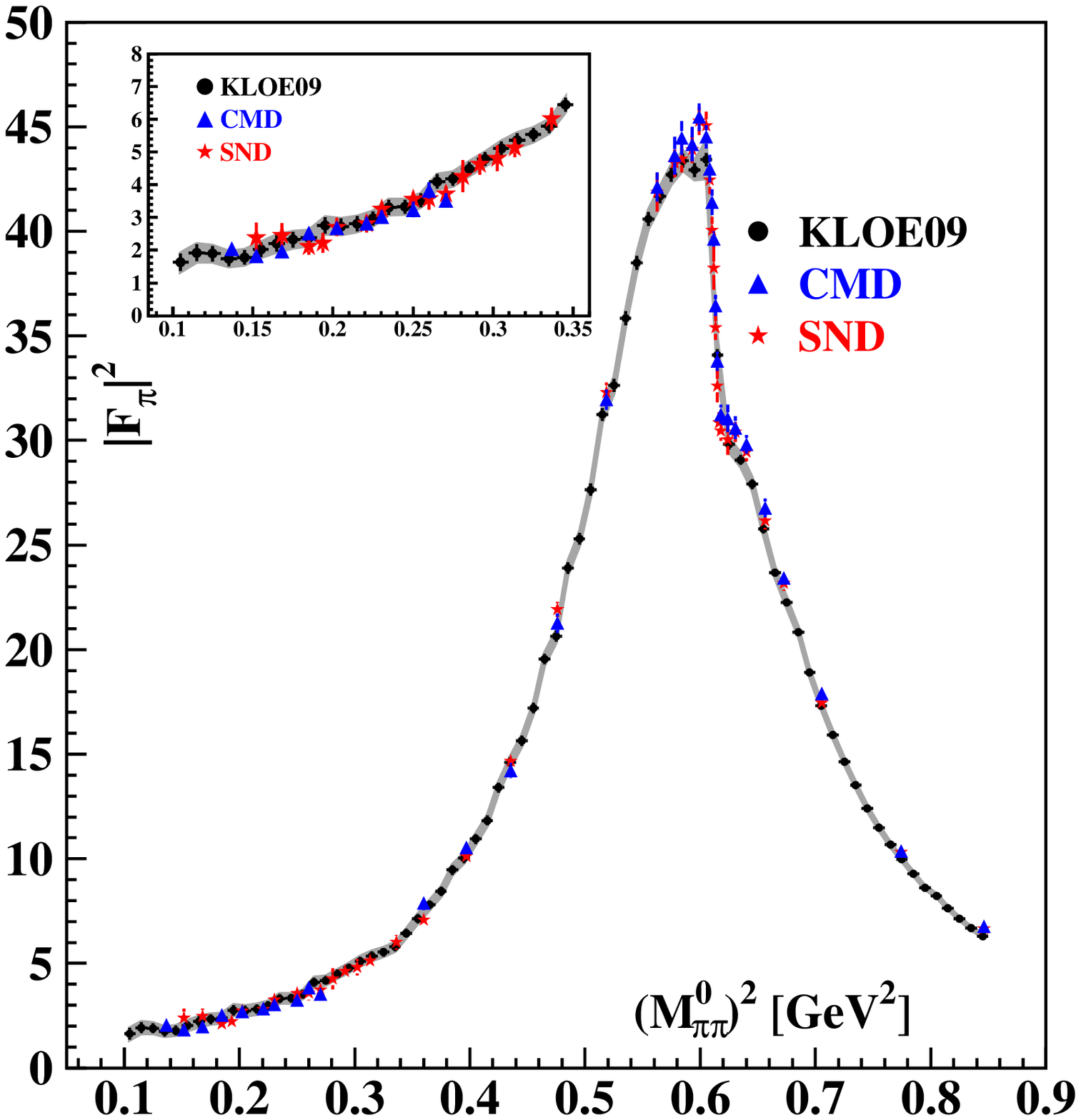}
\figcaption{\label{fig:6} Pion form factor $|F_\pi^2|$ obtained in the
  present analysis (KLOE09) and results from the CMD and SND experiments. KLOE09 data
  points have statistical error attached, the grey band gives the
  statistical and systematic uncertainty (added in quadrature). Errors
  on CMD2 and SND points contain the combined statistical and
  systematic uncertainty.} 
\end{center}
\begin{center}
\includegraphics[width=18.pc]{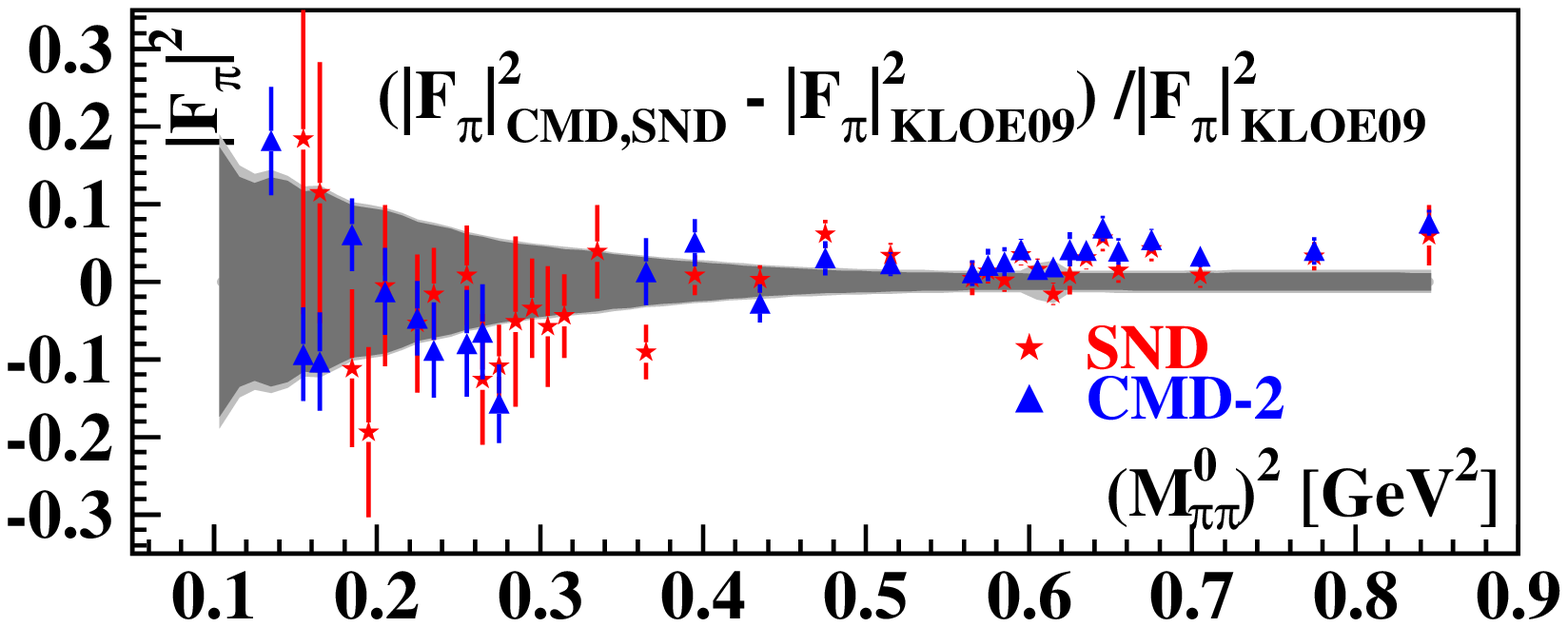}
\figcaption{\label{fig:7} Fractional difference between $|F_\pi^2|$
  from KLOE09 and the CMD and SND experiments. The band in dark grey
  represents the statistical error of the KLOE09 result, the band in
  lighter grey gives the statistical and systematic uncertainty (added
  in quadrature) for the KLOE09 result. Errors on CMD and SND points
  contain the combined statistical and systematic uncertainty.}
\end{center}
\subsection{Comparison with other experiments}
Fig.~\ref{fig:6} and Fig.~\ref{fig:7} show the KLOE09 result for $|F_\pi|^2$ together with
results from the CMD-2~\cite{CMD2,CMD2_2} and SND~\cite{SND} experiments
in Novosibirsk. While on the $\rho$-peak and above, the new result
confirms the KLOE08 result being lower than the Novosibirsk
results, below the $\rho$-peak the three experiments show good agreement.
\section{Forward-backward asymmetry}
The interference in the amplitudes for ISR and FSR is odd under the
exchange $\pi^+ \leftrightarrow \pi^- $. This gives rise to a
non-vanishing asymmetry of the distributions in the polar angle
$\theta$ for the pions~\cite{Binner99}. A common way to express this
is the {\it forward-backward} asymmetry $\mathcal{A}_{\rm FB}$: 
\begin{equation}
\label{eq:5}
  \mathcal{A}_{\rm FB}(M_{\pi\pi}^2) = \frac
  {N_{\pi^{+}}(\theta > 90^\circ)-N_{\pi^{+}}(\theta < 90^\circ)}
  {N_{\pi^{+}}(\theta > 90^\circ)+N_{\pi^{+}}(\theta < 90^\circ)}.
\end{equation}
This quantity is an ideal tool to test the validity of models used in
Monte Carlo to describe the pionic final state radiation. In a similar
way, radiative decays of the $\phi$ meson into scalars decaying into
$\pi^+\pi^-$ contribute to the asymmetry~\cite{Melnikov:2000gs,Czyz:2005}. As can be
seen in Fig.~\ref{fig:8} and Fig.~\ref{fig:9}, this has a large effect
on the asymmetry going from data taken at $\sqrt{s}\simeq 1$ GeV to
data taken at $\sqrt{s}=M_\phi$, especially in the energy region below
the $\rho$-meson mass. Outside the $\phi$-resonance, the
asymmetry is almost completely dominated by the pionic final state
radiation, while on the peak of the resonance, the decays of the
$\phi$-meson to $f_0\gamma$ and also $\rho\pi$ contribute
significantly. A comparison with  a Monte Carlo prediction using the
PHOKHARA event generator~\cite{OlgaS} with a model for $\phi$-decays and parameters
from~\cite{Ambrosino:2006hb}, together with a pointlike-pion
description for the pionic final state radiation, shows a good agreement
with the data for both sets of data. Qualitatively, the theoretical descriptions used to  model the different contributions in the simulation agree well with the data, although at 
low $M_{\pi\pi}^2$ the data statistics becomes poor and the data asymmetry points have large errors. In particular, the {\it off-peak} data in 
Fig.~\ref{fig:8} shows very good agreement above 0.35
GeV$^2$ with the pointlike-pion description for FSR. Further work is
in progress to determine the impact of different models for FSR on the
asymmetry, as well as to estimate higher order
effects~\cite{ivashyn2009}. In future, the larger dataset from
2004-2005, which is almost 10 times larger than the data shown in 
Fig.~\ref{fig:9}, may be used to determine with high precision the 
parameters of the $\phi$ decay contributions, in combination with the
results from the neutral channel $\phi \to \pi^0\pi^0\gamma$ and the 
assumption of isospin symmetry. This will then in turn allow to perform a
precise {\it on-peak} measurement of the pion form factor down to the
production threshold, using the full data sample of about 2 fb$^{-1}$
accumulated by the KLOE experiment. 

\begin{center}
\includegraphics[width=18.pc]{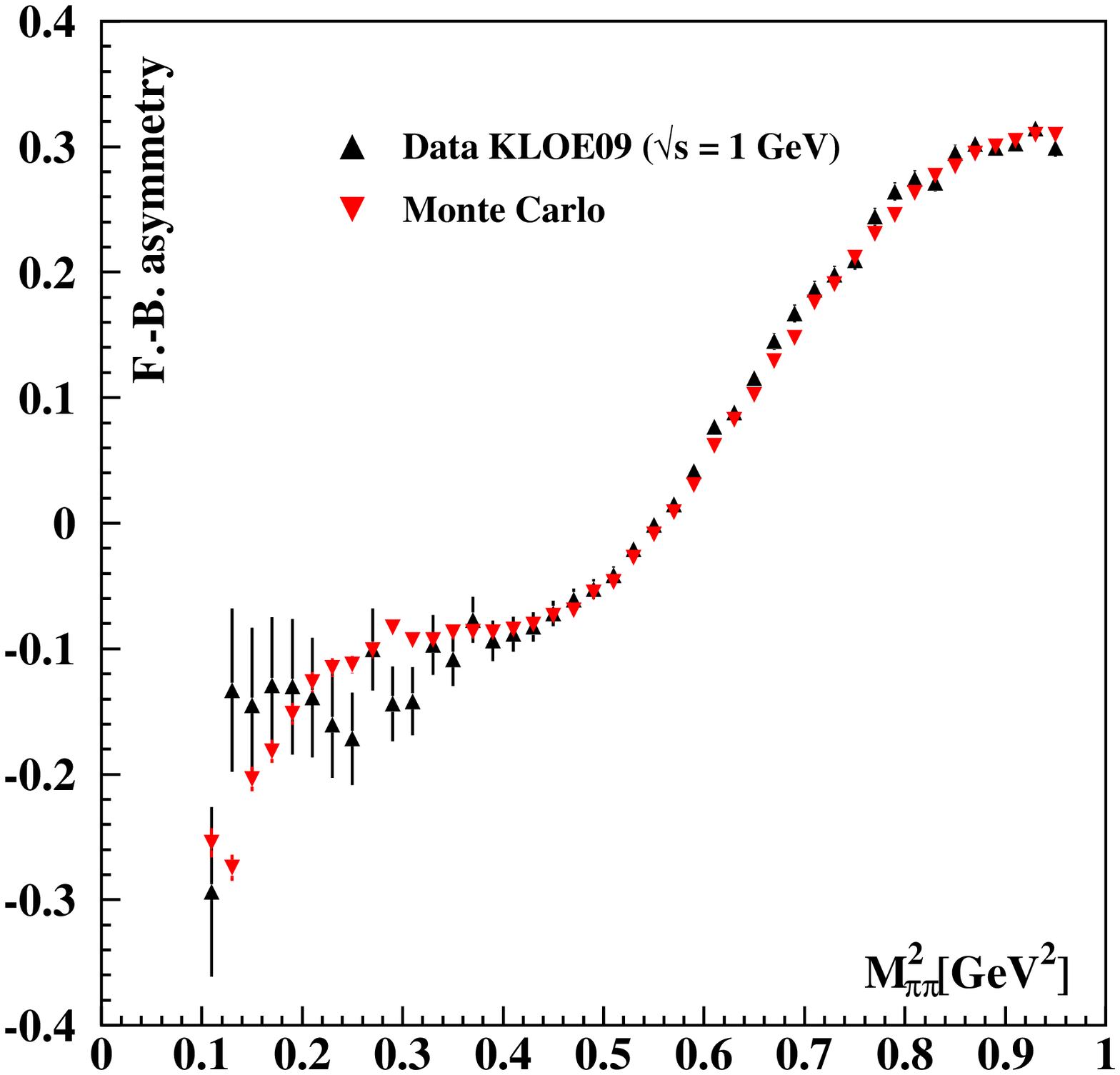}
\figcaption{\label{fig:8} Preliminary forward-backward asymmetry for
  KLOE09 data taken at $\sqrt{s}\simeq 1$ GeV, and the corresponding
  Monte Carlo prediction using the PHOKHARA event generator with model
  and parameters from~\cite{Ambrosino:2006hb}.} 
\end{center}
\begin{center}
\includegraphics[width=18.pc]{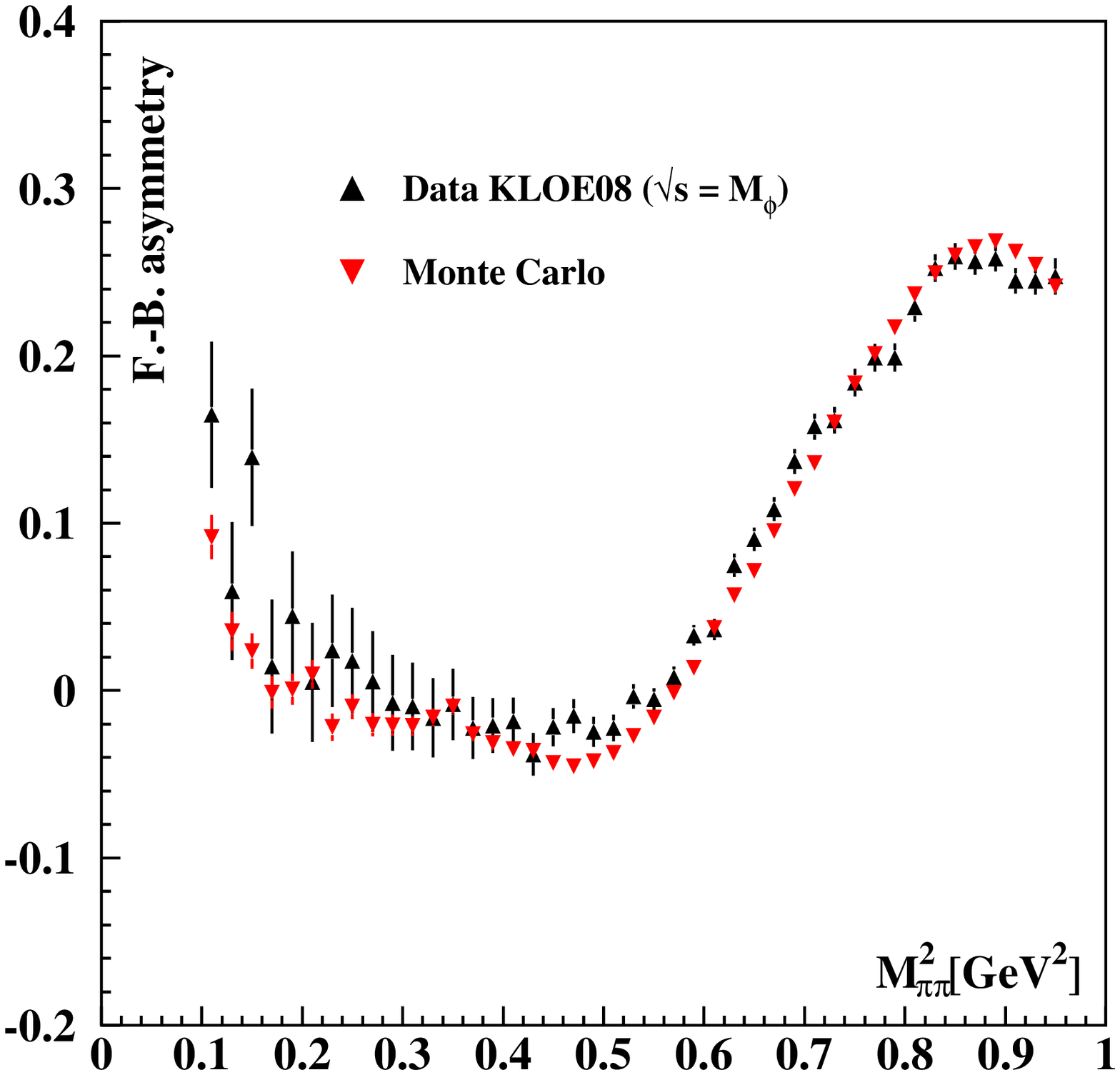}
\figcaption{\label{fig:9} Preliminary forward-backward asymmetry for
  KLOE08 data taken at $\sqrt{s}=M_\phi$, and the corresponding
  Monte Carlo prediction using the PHOKHARA event generator with model
  and parameters from~\cite{Ambrosino:2006hb}.} 
\end{center}

\section{Conclusions and outlook}
The KLOE experiment has performed a new measurement of
the pion form factor $|F_\pi|^2$ in the $M_{\pi\pi}^2$ range between 0.1 and 0.85
GeV$^2$. The result is in very good agreement with the previous KLOE
result, and extends it down to the two-pion threshold. Reasonable
agreement was found (especially at low energies) with the results
obtained from the Novosibirsk experiments CMD-2 and SND. The new KLOE
result further confirms the discrepancy between the
Standard Model evaluation for $a_\mu$ and the experimental value
measured by the (g-2) collaboration at BNL.\\

A next step at KLOE will be the measurement of the pion form factor
using a normalization to radiative muon events in each bin. In this
way, many theoretical uncertainties would become negligible, since the
radiator function, the vacuum polarisation and the absolute luminosity
would cancel out in the ratio of $\pi\pi\gamma$ over $\mu\mu\gamma$ events to first order. Pions and muons are 
separated and identified using kinematical variables (e.g. the 
aforementioned trackmass variable)~\cite{Muller:2006bk}. The analysis
is in a very advanced state and a systematic precision similar to the one obtained
in the absolute measurement is expected.

The forward-backward asymmetry $\mathcal{A}_{\rm FB}$ is an important
tool to test models for pionic final state radiation and radiative
decays of the $\phi$ mesons to scalars. A good check on the validity
of models and their parameters is crucial for precise measurements of
the pion form factor below 1 GeV using initial state radiation, especially when
running at the energy of $\sqrt{s}=M_\phi$.  

\end{multicols}

\vspace{-2mm}
\centerline{\rule{80mm}{0.1pt}}
\vspace{2mm}

\begin{multicols}{2}

\end{multicols}

\vspace{5mm}


\begin{thebibliography}{90}

\vspace{3mm}

\bibitem{Jegerlehner:2009ry} 
  F.~Jegerlehner and A.~Nyffeler,
  Phys.\ Rept.\  {\bf 477} (2009) 1

\bibitem{Teubner:2009} T.~Teubner, these proceedings

\bibitem{Davier:2009} M.~Davier, these proceedings

\bibitem{g-2:2006}
  G.~W.~Bennett {\it et al.} [Muon g-2 Coll.],
  Phys.\ Rev.\  D {\bf 73} (2006) 072003

\bibitem{Roberts:2009} B.~L.~Roberts, these proceedings

\bibitem{actis:2009gg}
  S.~Actis {\it et al.}, arXiv:0912.0749 [hep-ph]

\bibitem{plb606}
 A.~Aloisio {\it et al.} [KLOE Coll.], Phys. Lett. B {\bf 606} (2005) 12

\bibitem{plb670}
  F.~Ambrosino {\it et al.}  [KLOE Coll.],
  Phys.\ Lett.\  B {\bf 670}, 285 (2009) 

\bibitem{KLOE:DC}
M.~Adinolfi {\it et al.},
Nucl. Inst. Meth. A {\bf 488}, 51 (2002) 

\bibitem{KLOE:EMC} 
M.~Adinolfi {\it et al.},
Nucl. Inst. Meth. A {\bf 482}, 364 (2002) 

\bibitem{Binner99} 
S.~Binner, J.~H.~K\"uhn and K.~Melnikov,
  Phys.\ Lett.\  B {\bf 459}, 279 (1999)

\bibitem{Leone:2007zz}
D.~Leone, PhD thesis, KA-IEKP-2007-7 (2007)

\bibitem{Amsler:2008zzb}
C.~Amsler {\it et al.}  [Particle Data Group],  Phys.\ Lett.\  B {\bf 667}, 1 (2008)

\bibitem{CarloniCalame:2000pz}
  C.~M.~Carloni Calame {\it et al.},
  Nucl.\ Phys.\  B {\bf 584}, 459 (2000) 

\bibitem{Balossini:2006wc}
  G.~Balossini {\it et al.}, Nucl.\ Phys.\  B {\bf 758}, 227  (2006)

\bibitem{Ambrosino:2006te}
  F.~Ambrosino {\it et al.} [KLOE Coll.],
  Eur.\ Phys.\ J.\  C {\bf 47},  589 (2006)

\bibitem{Czyz:2003}
H.~Czy{\.z}, A.~Grzeli\'nska, J.~K\"uhn, G.~Rodrigo,
Eur. Phys. J. C {\bf 27}, 563 (2003)

\bibitem{Jeger_alpha}
  F.~Jegerlehner, Nucl.\ Phys.\ Proc.\ Suppl.\  {\bf 162}, 22 (2006)

\bibitem{Czyz:2005}
H.~Czy{\.z}, A.~Grzeli\'nska, J.~K\"uhn, 
Phys. Lett. B {\bf 611},  116 (2005)

\bibitem{Melnikov:2000gs} K.~Melnikov, F.~Nguyen, B.~Valeriani, G.~Venanzoni, 
{\it Phys. Lett. B} {\bf 477}, 114 (2000)

\bibitem{Brodsky:67}
S.~J.~Brodsky, E.~De~Rafael, Phys. Rev. {\bf 168}, 1620 (1967)

\bibitem{CMD2} 
  R. R. Akhmetshin et al. [CMD-2 Coll.], {\it Phys. Lett. B} {\bf 648}, 28 (2007)
  
\bibitem{CMD2_2} 
  R. R. Akhmetshin et al. [CMD-2 Coll.], {\it JETP Lett.} {\bf 84}, 413 (2006)

\bibitem{SND}
  M. N. Achasov et al. [SND Coll.], {\it J. Exp. Theor. Phys.} {\bf
    103}, 380 (2006)

\bibitem{OlgaS}
 O.~Shekhovtsova, PHOKHARA6.1\\
 \mbox{(\url{http://ific.uv.es/~rodrigo/phokhara/})}, unpublished  

\bibitem{Ambrosino:2006hb}
  F.~Ambrosino {\it et al.}  [KLOE Coll.],
  Eur.\ Phys.\ J.\  C {\bf 49},  473 (2007)

\bibitem{ivashyn2009}
S.~Ivashyn, H.~Czy\.z, A.~Korchin, arXiv:0910.5335 (2009)

\bibitem{Muller:2006bk}
 S.~E.~M{\" u}ller, F.~Nguyen, Nucl. Phys. Proc. Suppl. {\bf 162},
90 (2004)




\end{thebibliography}
\end{document}